\documentclass[draft,12pt]{article}
\usepackage{amsmath,amsfonts,palatino,amsthm,epsf}
\newcommand{\beq}{\begin{equation}}
\newcommand{\eeq}{\end{equation}}

\begin{document}

\title{Yang-Mills Interactions and Gravity in Terms of Clifford Algebra}
\author{Wei Lu
\thanks{New York, USA, email address: weiluphys@yahoo.com}
\\
\\
}
\maketitle

\begin{abstract}
A model of Yang-Mills interactions and gravity in terms of the Clifford algebra $C\!\ell_{0,6}$ is presented. The gravity and Yang-Mills actions are formulated as different order terms in a generalized action. The feebleness of gravity as well as the smallness of the cosmological constant and theta terms are discussed at the classical level. The invariance groups, including the de Sitter and the Pati-Salam $SU(4)$ subgroups, consist of gauge transformations from either side of an algebraic spinor. Upon symmetry breaking via the Higgs fields, the remaining symmetries are the Lorentz $SO(1,3)$, color $SU(3)$, electromagnetic $U(1)_{EM}$, and an additional $U(1)^\prime$.  The first generation leptons and quarks are identified with even and odd parts of spinor idempotent projections. There are still several shortcomings with the current model. Further research is needed to fully recover the standard model results. 
\end{abstract}

{\bf PACS numbers}. 12.10.Dm, 04.20.Cv, 11.15.Ex, 11.10.Ef.

{\bf Keywords}. Clifford algebra, Yang-Mills interactions, gravity.

\newpage

\section{Introduction}

The similarity of the gravitation field and the non-Abelian Yang-Mills field \cite{YANG} inspired the formulation of gravity as a gauge theory of the Lorentz group \cite{LORE}, the Poincar\'e group \cite{POIN}, or the (anti-)de Sitter group \cite{DESI}. The list can go on. However, the gravity and Yang-Mills actions are usually constructed in different ways. As said by C. N. Yang \cite{YANG2}, ``that gravitation is a gauge field is universally accepted, although exactly how it is a gauge field is a matter still to be clarified''. 

We present a model of Yang-Mills interactions and gravity in terms of the Clifford algebra $C\!\ell_{0,6}$ valued spinors, Higgs fields, and gauge fields on the four-dimensional spacetime manifold. Although the Yang-Mills and gravity fields are treated on the same footing as local gauge fields from the outset, they show different dynamics, due to symmetry breaking via the Higgs fields and non-degenerate vacuum expectation value (VEV) of the vierbein field. We subscribe to the notion that an infinite number of terms allowed by symmetry requirements should be included in the action for gauge fields. The gravity and Yang-Mills actions are formulated as different order terms. The key to make practical predictions is to recognize that they must be made within the context of separation of energy scales, so that only the first few terms of the action are relevant in the low-energy limit. 

The Clifford algebra, also known as the geometric algebra or the spacetime algebra for the specific case $C\!\ell_{1,3}$, is widely used as a computational tool in physics\cite{CA1,CA0,CA00,CA2,CA3,CA4}. One application is to formulate the Dirac electron as one of the two minimal left ideals of $C\!\ell_{1,3}$ \cite{CA1,CA2}, where ideals are defined as idempotent projections of an algebraic spinor. Hestenes \cite{WEAK} noticed that both of the minimal left ideals, regarded as electron and neutrino wave functions, respectively, are needed in a unified electroweak theory. Further approaches\cite{SM2, SM1}, in terms of seven- or higher-dimensional Clifford algebras, also make use of idempotent projections for the identification of different fermions and show how the standard model gauge symmetries arise from the Clifford algebras. 

With the observation that there are 16 Weyl spinors (including the right-handed neutrino) with $16\times4=64$ total real components in the first generation, we propose that the Clifford algebra $C\!\ell_{0,6}$, with $2^{6}=64$ degree of freedom, is the choice with minimal Clifford algebra dimensions for a model of Yang-Mills interactions and gravity. While the conventional Dirac matrix operators $\gamma_{1}, \gamma_{2}, \gamma_{3}$ correspond to vectors in $C\!\ell_{0,6}$, the matrix operator $\gamma_{0}$ corresponds to a trivector. This is different from all previous approaches\cite{CA1, WEAK, SM1, SM2}. The availability of the double-sided gauge transformations on the algebraic spinors \cite{SM2} and the identification of the operator $\gamma_{0}$ as a trivector enable a {\it nontrivial} integration of spacetime and other gauge symmetries in a six-dimensional Clifford algebra, which is otherwise impossible for the conventional approaches \cite{CM}. We show in this paper that symmetry breaking is essential for obtaining the residual gauge symmetries and the eventual identification of the electrons, neutrinos, and quarks with even and odd parts of spinor idempotent projections. 

This paper is structured as follows: Section 2 is a brief introduction to the Clifford algebra $C\!\ell_{0,6}$. In section 3, the algebraic spinor is defined. And gauge symmetries are discussed based on the invariance of a spinor bilinear. In section 4, we study the symmetry breaking of the gauge symmetries via Higgs fields. The first generation fermions are identified with idempotent projections of the algebraic spinor. Section 5 introduces the gauge field $1$-forms and gauge-covariant derivatives. We present the spinor action with both Dirac-type and Majorana-type (only for the right-handed neutrino) mass terms. Section 6 gives the definition of gauge strength curvature $2$-forms. The gravity and Yang-Mills actions are formulated as different order terms in a generalized action. In the last section we draw our conclusions and discuss the shortcomings of the current model. 

\section{The Clifford Algebra $C\!\ell_{0,6}$}

We begin with a brief introduction to the Clifford algebra $C\!\ell_{0,6}$. It is defined by the anticommutators of the {\it orthonormal vector basis} $\{ \Gamma_{j}, \gamma_{j}; {j}= 1,2,3\}$ 
\begin{align}
&\Gamma_{j}\Gamma_{k}+\Gamma_{k}\Gamma_{j}=-2\delta_{jk},\\
&\gamma_{j}\gamma_{k}+\gamma_{k}\gamma_{j}=-2\delta_{jk},\\
&\Gamma_{j}\gamma_{k}+\gamma_{k}\Gamma_{j}=0,
\end{align}
where ${j}, {k}=1, 2, 3.$ All the basis vectors are {\it spacelike}. There are ${{{{{{{{{{{{{{{{{{{{{\binom{6 }{k}}}}}}}}}}}}}}}}}}}}}} $ independent $k $-vectors. The complete basis for $C\!\ell_{0,6}$ is given by the set of all the $k $-vectors. A generic element in $C\!\ell_{0,6}$ is called a {\it multivector}. Any multivector can be
expressed as a linear combination of $2^{6}=64$ basis elements. 

The two trivectors
\begin{align}
&\gamma_{0} \equiv \Gamma_{1}\Gamma_{2}\Gamma_{3}, \\
&\Gamma_{0} \equiv \gamma_{1}\gamma_{2}\gamma_{3}
\end{align}
square to 1, so they are {\it timelike}. The orthonormal vector-trivector basis $\{\gamma_I, I= 0,1,2,3\}$ defines the spacetime Clifford algebra $C\!\ell_{1,3}$. The unit pseudoscalar
\begin{equation}
i\equiv \Gamma_{1}\Gamma_{2}\Gamma_{3}\gamma_1\gamma_2\gamma_3=\gamma_0\gamma_{1}\gamma_{2}\gamma_{3}=\gamma_0\Gamma_{0}
\end{equation}
squares to $-1$, anticommutes with odd grade elements, and commutes with even grade
elements. A multivector is said to be {\it even/odd} if it commutes/anticommutes with the
unit pseudoscalar.

The {\it reversion} of a multivector $M\in C\!\ell_{0,6}$, denoted $\tilde{M}$, reverses the
order in any product of vectors. There are algebraic properties $(MN)^{\tilde{}} = \tilde{N}\tilde{M}$ and $\left\langle MN\right\rangle
=\left\langle NM\right\rangle$ for any multivectors $M$ and $N$, where $\left\langle \cdots\right\rangle$ means the scalar part of the enclosed expression. The {\it length} of a multivector $M$ is defined as
\begin{equation}
|M| \equiv \sqrt{\left\langle M^{\dagger}M \right\rangle},
\end{equation}
where $M^{\dagger} \equiv i\tilde{M}(-i)$ is the {\it Hermitian conjugate}.

A projection operator squares to itself. The {\it idempotents} are a set of projection operators
\begin{align}
&p_{\pm{k}} \equiv \frac{1}{2}(1\pm {\gamma_0}{\Gamma_k}{\gamma_k}), \label{IDEM1}\\
&P_0 \equiv p_{+1}p_{+2}p_{+3} + p_{-1}p_{-2}p_{-3}, \label{IDEM2}\\
&P_1 \equiv p_{+1}p_{-2}p_{-3} + p_{-1}p_{+2}p_{+3}, \label{IDEM3}\\
&P_2 \equiv p_{-1}p_{+2}p_{-3} + p_{+1}p_{-2}p_{+3}, \label{IDEM4}\\
&P_3 \equiv p_{-1}p_{-2}p_{+3} + p_{+1}p_{+2}p_{-3}, \label{IDEM5}\\
&P_{\pm} \equiv \frac{1}{2}(1\pm {\Gamma_0}{\Gamma_3}), \label{IDEM6}%
\end{align}
where $ p_{+k} + p_{-k} = 1, P_0 + P_1 + P_2 + P_3 = 1, P_+ + P_- = 1, k = 1,2,3.$


\section{The Spinor and Gauge Symmetries}

In this section we introduce the {\it algebraic spinor} $\psi\in C\!\ell_{0,6}$. It is a multivector which obeys the transformation law
\begin{equation}
\psi\quad\rightarrow\quad  \mathbb{L}\psi{\mathbb{R}},
\end{equation}
where $\mathbb{L}$ and $\mathbb{R} \in C\!\ell_{0,6}$ are left and right gauge transformations. It has been considered in earlier approaches \cite{CA1, SM2} that different gauge transformations could be applied to the left or right side of an algebraic spinor. The spinor bilinear
\begin{equation}
\left\langle \tilde{\psi}\gamma_0\psi{e^{\beta i}} \right\rangle, \label{BILI}
\end{equation}
where $\beta$ is an arbitrary real angle, is invariant if
\begin{align}
&\tilde{\mathbb{L}}\gamma_0 \mathbb{L} = \gamma_0, \label{EQ1}\\
&\mathbb{R}{e^{\beta i}}\tilde{\mathbb{R}} = {e^{\beta i}}, \label{EQ2}
\end{align}
where we restrict our discussion to the gauge transformations continuously connected to the identity. The physical significance of the bilinear \eqref{BILI} will be clear in section 5, as all the terms in the fermion action are similar spinor bilinears. In spite of additional gauge-covariant derivatives and Higgs fields, these spinor bilinears have the same gauge transformation properties as the bilinear \eqref{BILI}. Thus the fermion action as a whole is invariant under the gauge transformations satisfying equations \eqref{EQ1}, \eqref{EQ2}. 

The general solution of these equations has the form
\begin{align}
&\mathbb{L} = e^{\frac{1}{2}\Theta_l}, \label{ROTO1}\\
&\mathbb{R} = e^{\frac{1}{2}\Theta_r}, \label{ROTO2}
\end{align}
where $\Theta_l$ is a linear combination of 28 gauge transformation generators
\begin{equation}
\{\gamma_I, \gamma_I\gamma_J, -\gamma_0\Gamma_j, \Gamma_0\Gamma_j, i\Gamma_j, \Gamma_0\Gamma_j\gamma_k ; j, k = 1,2,3, I, J = 0,1,2,3, I > J \}, \label{LROT}
\end{equation}
and $\Theta_r$ is a linear combination of 16 gauge transformation generators
\begin{equation}
\{i\} \oplus \{\gamma_0\Gamma_j, \Gamma_0\gamma_j, \Gamma_j\gamma_k; j, k = 1,2,3\}. \label{RROT}
\end{equation}
The de Sitter algebra $\{\gamma_I, \gamma_I\gamma_J\}$\footnote{See ref. \cite{ADS} for similar embedding of the anti-de Sitter algebra into $C\!\ell_{3,1}$.} and the weak interaction $su(2)$ algebra $\{-\gamma_0\Gamma_j\}$ are commuting subalgebras of the left gauge transformations. The $u(1)$ algebra $\{i\}$ and the Pati-Salam \cite{SU4} $su(4)$ algebra $\{\gamma_0\Gamma_j, \Gamma_0\gamma_j, \Gamma_j\gamma_k\}$ are commuting subalgebras of the right gauge transformations. It is well known that the Pati-Salam $SU(4)$ group is also a subgroup of $SO(10)$ in the $SO(10)$ grand unified theory \cite{SO10}. Here gauge transformation generators are multiplied by a factor of $1/2$, so that the structure constants of commutation relations are consistent with those of the conventional Lie algebra. For example, the weak gauge transformations are
\begin{equation}
\mathbb{L} = e^{\theta_1(\frac{1}{2}\Gamma_2\Gamma_3) + \theta_2(\frac{1}{2}\Gamma_3\Gamma_1)
				+ \theta_3(\frac{1}{2}\Gamma_1\Gamma_2)}.
\end{equation}

It will be shown in later sections that the local gauge fields of the de Sitter symmetry would give rise to gravity, while other gauge interactions are mediated by the local gauge fields of the rest of gauge transformations. It is remarkable that the left and right gauge groups contain both gravitational and internal gauge transformations. 

The {\it left-} and {\it right-handed} spinors correspond respectively to odd and even multivectors,
\begin{align}
&\psi = \psi_- + \psi_+, \\
&\psi_\mp \equiv \frac{1}{2}(\psi \pm i\psi i). 
\end{align}
We note that Hestenes proposed an opposite assignment for the left- and right-handed spinors \cite{WEAK}. Transformations $\{\gamma_I, i\Gamma_j, \Gamma_0\Gamma_j\gamma_k\}$ change the chirality of a spinor, while the other gauge transformations preserve chirality. The weak gauge transformations act on the left- and right-handed spinors in the same way, thus our model is left-right symmetric \cite{SU4, SO10}.

\section{The Higgs fields and Symmetry Breaking}

For breaking of the gauge symmetries, {\it Higgs fields}\footnote{We call them Higgs fields in the general sense that they are $0$-forms with symmetry-breaking VEVs, which are invariant under Lorentz gauge transformations \eqref{LO}.} $\phi, \phi_1, \phi_2, \phi_3,$ and $\Phi$ are introduced. They are multivectors which obey the transformation laws
\begin{align}
&\phi\quad\rightarrow\quad  {\mathbb{L}}\phi{\mathbb{L}}^{-1}, \\
&\phi_1\quad\rightarrow\quad  {\mathbb{L}}\phi_1{\mathbb{L}}^{-1}, \\
&\phi_2\quad\rightarrow\quad  {\mathbb{L}}\phi_2{\mathbb{L}}^{-1}, \\
&\phi_3\quad\rightarrow\quad  {\mathbb{L}}\phi_3{\mathbb{L}}^{-1}, \\
&\Phi\quad\rightarrow\quad  {\mathbb{R}}^{-1}\Phi{\mathbb{R}}.
\end{align}
It's easy to check that, besides the spinor bilinear \eqref{BILI}, additional spinor bilinears
\begin{align}
&\left\langle \tilde{\psi}\gamma_0 \phi_1 \phi_3(\psi - \phi
\psi i)\Phi \right\rangle, \label{BILII1}\\
&\left\langle \tilde{\psi}\gamma_0 \phi_2 \phi_3(\psi - \phi
\psi i)\Phi \right\rangle, \label{BILII2}\\
&\left\langle \tilde{\psi}\gamma_0 \phi_3 (\psi - \phi
\psi i)(1 + c_1 \Phi^2) \right\rangle, \label{BILII3}\\
&\left\langle \tilde{\psi}\gamma_0 \phi_3 (\psi + \phi
\psi i)(1 + c_2 \Phi^2) \right\rangle, \label{BILII4}
\end{align}
are invariant under the gauge transformations \eqref{ROTO1} \eqref{ROTO2}, where $c_1$ and $c_2$ are real constants. But how the symmetry is realized in nature depends on the properties of the vacuum state. When a Higgs field acquires a VEV, which is not invariant under the original gauge transformations, the symmetry is said to be broken. It is out of scope of this paper to study the self-interacting potentials and dynamics of the Higgs sector. The Higgs fields are just {\it assumed} to have VEVs as
\begin{align}
&\bar{\phi}=i, \label{PHI}\\
&\bar{\phi}_1=v_1\Gamma_2\Gamma_3, \label{PHI1}\\
&\bar{\phi}_2=v_2\Gamma_3\Gamma_1, \label{PHI2}\\
&\bar{\phi}_3=P_-, \label{PHI3}\\
&\bar{\Phi}=V\Gamma_0 P_0, \label{BPHI}
\end{align}
where $P_0, P_-$ are idempotents \eqref{IDEM2} \eqref{IDEM6}, $\bar{\phi}$ and $\bar{\phi}_3$ are of fixed lengths, $v_1$, $v_2$, and $V$ are real constants which modulate the lengths of the VEVs. With the substitution of Higgs fields by their VEVs, the bilinears \eqref{BILII1} \eqref{BILII2} \eqref{BILII3} \eqref{BILII4} are rewritten as
\begin{align}
&\left\langle \tilde{\psi}_+\gamma_0 (v_1\Gamma_2\Gamma_3 P_-)\psi_+(V\Gamma_0 P_0) \right\rangle, \\
&\left\langle \tilde{\psi}_+\gamma_0 (v_2\Gamma_3\Gamma_1 P_-)\psi_+(V\Gamma_0 P_0) \right\rangle, \\
&\left\langle \tilde{\psi}_-\gamma_0 P_- \psi_+ (1 + c_1 V^2P_0) \right\rangle, \\
&\left\langle \tilde{\psi}_+\gamma_0 P_- \psi_- (1 + c_2 V^2P_0) \right\rangle, 
\end{align}
and the parity symmetry is maximally broken. In the next section, terms related to the bilinears \eqref{BILII1}, \eqref{BILII2}, \eqref{BILII3}, and \eqref{BILII4} are included in the spinor action as mass terms. After symmetry breaking, the residual symmetries of these bilinears are the Lorentz $SO(1,3)$ gauge transformations 
\begin{equation}
\{\gamma_I\gamma_J; I, J = 0,1,2,3, I > J \}, \quad \text{for}\, \mathbb{L} = e^{\frac{1}{2}\Theta_l}, \label{LO}
\end{equation}
the color $SU(3)$ gauge transformations \footnote{The $SU(3)$ gauge transformations may not work for antiparticles. This needs to be addressed in further research.} 
\begin{equation}
\left\{
\begin{array}{rl}
&\frac{1}{2}(\Gamma_1\Gamma_2 + \gamma_1\gamma_2), 
\frac{1}{2}(\Gamma_1\gamma_2 - \gamma_1\Gamma_2), 
\frac{1}{2}(\Gamma_1\gamma_1 - \Gamma_2\gamma_2), \\
&\frac{1}{2}(\Gamma_3\Gamma_1 + \gamma_3\gamma_1), 
\frac{1}{2}(\Gamma_3\gamma_1 - \gamma_3\Gamma_1), \\
&\frac{1}{2}(\Gamma_2\Gamma_3 + \gamma_2\gamma_3), 
\frac{1}{2}(\Gamma_2\gamma_3 - \gamma_2\Gamma_3), \\
&\frac{1}{2\sqrt{3}}(\Gamma_1\gamma_1 + \Gamma_2\gamma_2 - 2\Gamma_3\gamma_3), 
\end{array}
\right\} \quad \text{for}\, \mathbb{R} = e^{\frac{1}{2}\Theta_r}, 
\label{SU3}
\end{equation}
the electromagnetic $U(1)_{EM}$ synchronized double-sided gauge transformation
\begin{equation}
\left\{
\begin{array}{rl}
&\mathbb{L} = e^{\frac{1}{2}\theta \Gamma_1\Gamma_2}, \\
&\mathbb{R} = e^{-\frac{1}{2}\theta(\cos{\theta_0})^2 \Upsilon -\frac{1}{2}\theta (\sin{\theta_0})^2 i},
\end{array}
\right.
\label{EM}
\end{equation}
and an additional $U(1)^{'}$ gauge transformation
\begin{equation}
\mathbb{R} = e^{-\frac{3}{4}\theta^{'}(\Upsilon - i) }, \label{U1}
\end{equation}
where $\frac{1}{2}\Gamma_1\Gamma_2$ in the electromagnetic gauge transformation \eqref{EM} is the third component of the weak gauge algebra, the transformation generator $\Upsilon \equiv \frac{1}{3}(\Gamma_1\gamma_1 + \Gamma_2\gamma_2 + \Gamma_3\gamma_3)$ commutes with the color $su(3)$ generators, and $\tan{\theta_0}$ is related to the ratio of coupling constants of the $U(1)$ and $SU(4)$ gauge interaction. It is known that the unitary algebra $u(3)$ is embedded \cite{EMBE} in the orthogonal algebra $so(6) \simeq su(4)$. Removing $\Upsilon$ from $u(3)$ defines the color algebra $su(3)$. 

It should be noted that in the context of the Clifford algebra $C\!\ell_{7,0}$, Trayling and Baylis \cite{SM2} first pointed out that the weak $SU(2)$, color $SU(3)$, and electromagnetic $U(1)_{EM}$ symmetries are related to left-sided, right-sided, and synchronized double-sided gauge transformations, respectively. We also note that, in terms of the spacetime algebra $C\!\ell_{1,3}$, the local Lorentz symmetries were recognized very early on as left-sided gauge transformations \cite{CA1} in addition to the right-sided internal gauge transformations, and further exploited in the gauge theory of gravity \cite{GG}.

Now we are ready to identify idempotent projections of the left- and right-handed spinors
\begin{equation}
\psi = (P_- + P_+)(\psi_- + \psi_+)(P_0 + P_1 + P_2 + P_3)
\end{equation}
with the left-handed leptons, red, green, and blue quarks
\begin{equation}
\left\{
\begin{array}{rl}
&\nu_l \equiv P_+\psi_-P_0, \\
&e_l \equiv P_-\psi_-P_0, \\
&u_l \equiv P_+\psi_-P_1, \quad P_+\psi_-P_2, \quad \text{and} \quad P_+\psi_-P_3,  \\
&d_l \equiv P_-\psi_-P_1, \quad P_-\psi_-P_2, \quad \text{and} \quad P_-\psi_-P_3, 
\end{array}
\right.
\end{equation}
and the right-handed leptons, red, green, and blue quarks
\begin{equation}
\left\{
\begin{array}{rl}
&\nu_r \equiv P_-\psi_+P_0, \\
&e_r \equiv P_+\psi_+P_0, \\
&u_r \equiv P_-\psi_+P_1, \quad P_-\psi_+P_2, \quad \text{and} \quad P_-\psi_+P_3,  \\
&d_r \equiv P_+\psi_+P_1, \quad P_+\psi_+P_2, \quad \text{and} \quad P_+\psi_+P_3, 
\end{array}
\right.
\end{equation}
in the first generation. Because the product of $P_0$ with any generator in the color algebra $su(3)$ is zero, leptons are invariant under $su(3)$ color gauge transformations, while they could be regarded as the fourth color of $su(4)$ \cite{SU4}. For the electromagnetic gauge transformation \eqref{EM}, the electric charges\footnote{The electric charges $q_a$ are defined as $\mathbb{L}(\theta)\psi_a\mathbb{R}(\theta) = \psi_a exp(q_a\theta i)$, where $\psi_a$ is one of the idempotent projections of the left- or right-handed spinor, $\mathbb{L}$ and $\mathbb{R}$ are defined in \eqref{EM}. The trick of the calculation is that the transformation generators are equivalent to the pseudoscalar $i$ with different factors for different $\psi_a$. The pseudoscalar $i$ acquires a minus sign while moving from the left to the right of $\psi_a$, if $\psi_a$ is an odd multivector. We note that symmetry breaking is essential for obtaining the residual electromagnetic gauge symmetry and the eventual identification of the electrons, neutrinos, and quarks with even and odd parts of spinor idempotent projections. The symmetry-breaking VEVs of the Higgs fields (which involve the idempotents $P_0$ and $P_-$) are the causes. The spinor idempotent projections and the calculated electric charges are the effects.} $q_a$ are calculated as $0, -1, \frac{1}{2} + \frac{1}{6}(\cos{\theta_0})^2 - \frac{1}{2}(\sin{\theta_0})^2,$ and $
-\frac{1}{2} + \frac{1}{6}(\cos{\theta_0})^2 - \frac{1}{2}(\sin{\theta_0})^2$ 
for neutrino, electron, up quarks, and down quarks, respectively. For the $U(1)^{'}$ gauge transformation \eqref{U1}, the charges\footnote{The charges $q_a^{\prime}$ are defined as $\psi_a\mathbb{R}(\theta^{\prime}) = \psi_a exp(q_a^{\prime}\theta^{\prime} i)$, where $\mathbb{R}$ is defined in \eqref{U1}.} $q_a^{\prime}$ of the {\it fifth force} are calculated as $0$ and $1$ for leptons and quarks, respectively. If the coupling constant of the fifth force is set to zero by an unknown mechanism (which means $\theta_0 = 0$), the standard model electric charge assignments are recovered. For example, one could simply assume that the $U(1)$ transformation in \eqref{RROT} should be left as a global gauge symmetry, instead of being locally gauged. However, we don't rule out the possibility that the angle $\theta_0$ may be extremely small but not zero. The very weak fifth force and the minute modifications to the electric charges may be revealed if one looks more carefully into cosmological observations, especially in the dark sector (dark matter and dark energy).

Those who are versed in conventional grand unified theories would point out that the breaking of symmetries should be considered in (at least) two steps with different energy scales. For example, the first step breaks the symmetries down to $SO(1,3)\otimes SU(3)\otimes SU(2)_L \otimes U(1)_Y \otimes U(1)^{'}$. The second step breaks the symmetries down further to $SO(1,3)\otimes SU(3)\otimes U(1)_{EM} \otimes U(1)^{'}$. The conventional left-right symmetric models contain $SU(2)_L \otimes SU(2)_R$ along with a Higgs mechanism for breaking the $SU(2)_R$ symmetry and retaining $SU(2)_L$ symmetry at the intermediate electroweak energy scale. It seems impossible that a similar mechanism could be employed in our approach, since there are only three degrees of freedom for the weak transformations applying to both the left- and right-handed components of the spinor. We will revisit this issue in next section.

\section{The Gauge Fields and Spinor Action}
Before presenting the spinor action, we first introduce the gauge field $1$-forms and gauge-covariant derivatives on the spacetime four-dimensional manifold. The coordinates are written as $x = (x_{\mu})$, where $\mu = 0, 1, 2, 3$. The local gauge transformations are $\mathbb{L}(x)$ and $\mathbb{R}(x)$ \eqref{ROTO1}, \eqref{ROTO2}, thus the gauge transformation laws of the spinor and Higgs fields are coordinate-dependent. The {\it gauge fields} are connection 1-forms
\begin{align}
&a(x)\equiv a_{\mu}(x)dx^{\mu}, \\
&A_{u(1)}(x)\equiv A_{u(1)\mu}(x)dx^{\mu}, \\
&A_{su(4)}(x)\equiv A_{su(4)\mu}(x)dx^{\mu},
\end{align}
valued in the left gauge transformation algebra \eqref{LROT} and right gauge transformation algebras $u(1)\oplus su(4)$ \eqref{RROT}, respectively. Here we adopt the summation convention for repeated indices. The gauge fields obey the local gauge transformation laws
\begin{align}
&a(x)\quad\rightarrow\quad  {\mathbb{L}(x)}a(x){\mathbb{L}(x)}^{-1} - d{\mathbb{L}(x)}{\mathbb{L}(x)}^{-1}, \\
&A_{u(1)}(x)\quad\rightarrow\quad  {\mathbb{R}_{u(1)}(x)^{-1}A_{u(1)}(x){\mathbb{R}_{u(1)}(x)}} + {\mathbb{R}_{u(1)}(x)^{-1}d{\mathbb{R}_{u(1)}(x)}}, \\
&A_{su(4)}(x)\quad\rightarrow\quad  {\mathbb{R}_{su(4)}(x)^{-1}A_{su(4)}(x){\mathbb{R}_{su(4)}(x)}} + {\mathbb{R}_{su(4)}(x)^{-1}d{\mathbb{R}_{su(4)}(x)}}. 
\end{align}
The {\it spin connection} $1$-form $\omega$ valued in the Lorentz algebra, the  {\it weak connection} $1$-form $W$ valued in weak $su(2)$ algebra, and the  $1$-form $(1/l)e$ (where $l$ is a constant of the dimensions of length) valued in $\{\gamma_I; I = 0, 1, 2, 3\}$ are part of even and odd multivector gauge field $a = a_+ + a_-$, respectively. The {\it vierbein} $e$ gives rise to the {\it metric} $g_{\mu\nu} = \left\langle e_{\mu}e_{\nu} \right\rangle$. 
 
The gauge-covariant derivatives of the spinor field $\psi(x)$ and the Higgs field $\phi(x)$ are defined by
\begin{align}
&D\psi(x) \equiv d\psi(x) + a(x)\psi(x) - \psi(x) (A_{u(1)}(x) + A_{su(4)}(x)), \\
&D\phi(x) \equiv d\phi(x) + a(x)\phi(x) - \phi(x) a(x).
\end{align}
Here the connection fields are defined to absorb the coupling constants. With the replacement of the Higgs field $\phi$ by its VEV, $D\phi$ reduces to $2a_- i$, where gauge field $a_-$ is the odd multivector part of the gauge field $a$. The gauge- and diffeomorphism-invariant action for the spinor field is now written down as
\begin{equation}
S_F \sim \int{\left\langle \tilde{\psi}\gamma_0{e^{\alpha \phi}}(D\phi\wedge D\phi\wedge D\phi\wedge D\psi){e^{\beta i}} \right\rangle},
\end{equation}
where factors $e^{\alpha \phi}$ and $e^{\beta i}$ moderate the non-minimal coupling, which was shown to lead to parity violation effects \cite{NONM}. Similar fermion action (without factors $e^{\alpha \phi}$ and $e^{\beta i}$) was proposed by Pagels \cite{FERM}. The covariant derivative of the Higgs field plays a significant role in the construction of the spinor action. As opposed to the conventional approaches, the Higgs fields are integral part of the kinetic term in the spinor action. It can be checked that the spinor fields and the Higgs fields are dimensionless in this formulation of the spinor action. The conventional dimensions of the spinor fields are recovered if the factor of $1/l^3$ from $D\phi\wedge D\phi\wedge D\phi$ is absorbed by the rescaled spinor fields.

At the end of last section, we raised the issue about how to realize the chiral weak interactions in the current framework. In this regard, one could subtract the following terms
\begin{align}
&\int{\left\langle ((1-\phi_3)(\psi - \phi\psi i))\tilde{} \ \gamma_0{e^{\alpha \phi}}(D\phi\wedge D\phi\wedge D\phi\wedge D(\phi_3(\psi - \phi\psi i))){e^{\beta i}} \right\rangle}, \label{CHI1} \\
&\int{\left\langle (\phi_3(\psi - \phi\psi i))\tilde{} \ \gamma_0{e^{\alpha \phi}}(D\phi\wedge D\phi\wedge D\phi\wedge D((1-\phi_3)(\psi - \phi\psi i))){e^{\beta i}} \right\rangle}, \label{CHI2}
\end{align}
from the fermion action to prohibit interactions between right-handed fermions via the $W^+$ and $W^-$ weak gauge fields. More terms (not shown here) could be added to moderate the weak interactions of the right-handed neutral currents. However, we would not pursue this rather {\it ad hoc} procedure any further in this paper. 

While gauge field $a_-$ in $D\psi$ manifests itself in the spinor action as a universal Dirac-type mass contribution to fermions \cite{MASS}, other mass terms can be added to the spinor action, such as the Majorana-type mass terms for the right-handed neutrino
\begin{align}
&\int{\left\langle \tilde{\psi}\gamma_0{e^{\alpha_1 \phi}} \phi_1 \phi_3 (D\phi\wedge D\phi\wedge D\phi\wedge D\phi
)(\psi - \phi
\psi i)\Phi{e^{\beta_1 i}} \right\rangle}, \\
&\int{\left\langle \tilde{\psi}\gamma_0{e^{\alpha_2 \phi}} \phi_2 \phi_3 (D\phi\wedge D\phi\wedge D\phi\wedge D\phi
)(\psi - \phi
\psi i)\Phi{e^{\beta_2 i}} \right\rangle}, 
\end{align}
and the Dirac-type mass terms
\begin{align}
&\int{\left\langle \tilde{\psi}\gamma_0{e^{\alpha_3 \phi}} \phi_3 (D\phi\wedge D\phi\wedge D\phi\wedge D\phi)(\psi - \phi
\psi i)
(1 + c_1\Phi^2) e^{\beta_3 i} \right\rangle}, \\
&\int{\left\langle \tilde{\psi}\gamma_0{e^{\alpha_4 \phi}} \phi_3 (D\phi\wedge D\phi\wedge D\phi\wedge D\phi)(\psi + \phi
\psi i)
(1 + c_2\Phi^2) e^{\beta_4 i} \right\rangle}. 
\end{align}
The coefficient and phase parameters of each mass term can be adjusted for different fermion masses. If the VEVs of $\phi_1$ and $\phi_2$ become very large, the Majorana-type mass is much heavier than the Dirac-type mass. Thus a tiny effective mass is generated for the left-handed neutrino, known as the seesaw mechanism \cite{NEUT}. 

The charge conjugation of the spinor field $\psi$ is defined by
\begin{equation}
\psi^c \equiv e^{\zeta_1\Gamma_1\Gamma_2}\Gamma_1\Gamma_0\psi\Gamma_0e^{\zeta_2i}, \label{CH}
\end{equation}
where $e^{\zeta_1\Gamma_1\Gamma_2}$ and $e^{\zeta_2i}$ are arbitrary phase factors. The charge conjugation satisfies the properties $(\psi^c)^c = \psi$, $(\psi i)^c = - \psi^c i$, and $(\gamma_{I}\psi)^c = - \gamma_{I}\psi^c$. With the substitution of the Higgs fields by their VEVs, the Majorana-type mass terms can be shown to involve the right-handed neutrino $\nu_r$ and its charge conjugation $\nu_r^c$ with appropriate phase factors.

Since the gauge transformation laws of the covariant derivatives of the spinor field $D\psi(x)$ and the Higgs field $D\phi(x)$ are the same as the original spinor field and Higgs field, it can be readily checked that the gauge transformation properties of all the terms in the spinor action are the same as the bilinear \eqref{BILI}. Thus the spinor action as a whole is invariant under gauge transforms \eqref{ROTO1}, \eqref{ROTO2}. The spinor action is also diffeomorphism invariant, because it is the integral of $4$-forms on a $4$-manifold. We note that the Majorana-type mass terms, which correspond to bilinears \eqref{BILII1} and \eqref{BILII2} in last section, are essential for determining the residual symmetries after symmetry breaking.

The spinor action in flat spacetime can be obtained by substituting the Higgs fields, the spin connection $\omega$, and the gauge field $a_-$ with their VEVs \eqref{PHI}, \eqref{PHI1}, \eqref{PHI2}, \eqref{PHI3}, \eqref{BPHI}, $\bar{\omega} = 0$, and $\bar{a}_- = (1/l)\bar{e} = (1/l)\delta^{I}_{\mu}\gamma_I dx^{\mu} = (1/l)\gamma_{\mu} dx^{\mu}$, respectively. These VEVs, in particular the soldering form $\delta^{I}_{\mu}\gamma_I dx^{\mu}$, leave the originally local de Sitter gauge- and diffeomorphism-invariant action with a residual {\it global spacetime Lorentz symmetry}. 

For example, the part of the spinor action for the free electron in flat spacetime reduces to
\begin{equation}
S_{e} = \int {\left\langle \bar{\psi_e} \gamma^{\mu} \partial_{\mu} \psi_e i - m_e \bar{\psi_e} \psi_e \right\rangle} d^{4}x, \label{ELEC}
\end{equation}
where $\{\gamma^{\mu}\}$ is the {\it reciprocal frame} of $\{\gamma_{\mu}\}$ defined by $\left\langle \gamma^{\mu}\gamma_{\nu}\right\rangle = \delta^{\mu}_{\nu}$,  $\psi_e \equiv e_l + e_r = P_-\psi_-P_0 + P_+\psi_+P_0$, $\bar{\psi_e} \equiv {\psi^{\dagger}_e}\gamma_0 = i\tilde{\psi_e}(-i)\gamma_0$, and $m_e$ is the bare Dirac-type mass of the electron. The additional phase factors related to the pseudoscalar $i$ in the original spinor action have been eliminated after appropriate global phase transformations and rescaling for $\psi_e$. Here we have assumed that the integration over total derivatives could be discarded. We note that the even and odd parts of the electron spinor are rescaled by different factors. For general curved spacetime with torsion, these additional phase factors can not usually be eliminated. 

Variation with respect to the spinor field produces the Dirac equation in the form
\begin{equation}
\gamma^{\mu} \partial_{\mu} \psi_e i - m_e \psi_e = 0.
\end{equation}
If the electromagnetic interaction is included, one can show that the charge conjugation $\psi_e^c$ defined by \eqref{CH} satisfies the Dirac equation with an opposite electric charge, because of the properties $(\gamma^{\mu}\psi_e i)^c = \gamma^{\mu}\psi_e^c i$ and $(\gamma^{\mu}\psi_e)^c = - \gamma^{\mu}\psi_e^c$.

The electron action \eqref{ELEC} resembles the conventional action for the free Dirac field. Actually a map can be constructed by placing the Dirac column spinor $\hat{\psi}_e$ in one-to-one correspondence with the algebraic spinor $\psi_e$ via
\begin{equation}
\hat{\psi}_e = 
\left(
\begin{array}{rl}
a_1 + b_1\hat{i} \\
a_2 + b_2\hat{i} \\
a_3 + b_3\hat{i}  \\
a_4 + b_4\hat{i} 
\end{array}
\right)
\leftrightarrow \psi_e = w_n(a_n + b_n i),
\end{equation}
where $n = 1, 2, 3, 4$, $\hat{i}$ is the conventional unit imaginary number, $a_n$ and $b_n$ are real numbers, $w_1 \equiv 2\sqrt{2}P_-\gamma_0\gamma_1\gamma_0P_0$, $w_2 \equiv 2\sqrt{2}P_-\gamma_0P_0$, $w_3 \equiv 2\sqrt{2}P_+\gamma_1\gamma_0P_0$, and $w_4 \equiv 2\sqrt{2}P_+P_0$. Here the set of $\{w_n\}$ forms an orthonormal basis satisfying $\left\langle w^\dagger_n w_m \right\rangle = \delta_{nm}$. The mappings for the operators are
\begin{align}
\hat{\gamma}^\mu\hat{\psi}_e  &\leftrightarrow  \gamma^\mu\psi_e, (\mu =0,1,2,3)\\
\hat{i}\hat{\psi}_e  &\leftrightarrow  \psi_e i, \\
\hat{\gamma}^5\hat{\psi}_e  &\leftrightarrow  (-i)\psi_e i,
\end{align}
where $\hat{\gamma}^\mu$ and $\hat{\gamma}^5$ are the Dirac matrix operators in the Weyl representation \cite{PS}. And finally the mapping for the inner product is
\begin{align}
Re(\hat{\psi}_{e1}^\dagger\hat{\psi}_{e2}) &\leftrightarrow \left\langle \psi_{e1}^\dagger\psi_{e2} \right\rangle,\\
Im(\hat{\psi}_{e1}^\dagger\hat{\psi}_{e2}) &\leftrightarrow \left\langle \psi_{e1}^\dagger\psi_{e2} (-i) \right\rangle.
\end{align}
Armed with above mappings, we can check that the electron action \eqref{ELEC} indeed corresponds to the conventional action with the column spinors and the Dirac matrix operators. The spinor bilinear of the mass term in the action \eqref{ELEC} is a Lorentz scalar in the conventional parlance. And for that matter, all the bilinear covariants (Dirac current vectors, bivectors, etc.) can be easily translated back and forth between the conventional and the Clifford algebra formulations with the help of above mappings.

One can go even further, and find the column representations for all the leptons and quarks. We will not go into the details of further mappings in this paper. Suffice it to mention that the spinor columns for different fermions can be juxtaposed with each other to form a three dimensional matrix, with the de Sitter, weak, and right-sided gauge transformations acting as operator matrices on different abstract dimensions of the spinor matrix. However, for the rest of the left-sided gauge transformations which do not commute with the de Sitter and weak algebras, one can not translate them into matrix representations along the above line. 

\section{The Curvature and Gauge Action}

In this section, the Einstein-Cartan action and the Yang-Mills action are shown to be two relevant terms in a generalized action. We subscribe to the notion that an infinite number of terms allowed by symmetry requirements should be included in the action. The key to make practical predictions is to recognize that they must be made within the context of separation of energy scales, so that only the first few terms of the action are relevant in the low-energy limit.  The same notion, in the form of the {\it effective field theory}, has been applied to non-renormalizable as well as renormalizable quantum field theories \cite{EFFE}. 

It should be noted that a different approach by Smolin \cite{PLEB}, based on an extension of the Plebanski action to a Lie group containing the local Lorentz group, reached similar conclusions that the gravity and Yang-Mills actions are realized as different order terms in a unified theory.

We begin by introducing the {\it gauge strength curvature} $2$-forms by applying the covariant derivative to the $0$-form spinor $\psi$ and then to the $1$-form spinor $D\psi$
\begin{equation}
D(D\psi) = f\psi - \psi (F_{u(1)} + F_{su(4)}),
\end{equation}
where 
\begin{align}
&f \equiv da + a \wedge a, \\
&F_{u(1)} \equiv dA_{u(1)}, \\
&F_{su(4)} \equiv dA_{su(4)} + A_{su(4)} \wedge A_{su(4)}. 
\end{align}
Here the gauge-covariant derivative of an $n$-form spinor field $\Psi(x)$ is given by
\begin{equation}
D\Psi(x) \equiv d\Psi(x) + a(x) \wedge \Psi(x) - (-1)^n \Psi(x) \wedge (A_{u(1)}(x) + A_{su(4)}(x)). \\
\end{equation}
The generalized formal curvature $2$-form is defined as
\begin{equation}
\mathcal{F} \equiv \frac{1}{g}e^{\vartheta\phi}f + \frac{1}{g^\prime}e^{\vartheta^\prime i}F_{u(1)} + \frac{1}{g^{\prime\prime}}e^{\vartheta^{\prime\prime}i}F_{su(4)},
\end{equation}
where $g, g^\prime, g^{\prime\prime}, \vartheta, \vartheta^\prime$, and $\vartheta^{\prime\prime}$ are dimensionless constants. Because curvature 2-form $f$ includes both gravity and the weak interactions, they share the same bare coupling constant $g$. The observed feebleness of gravity will be discussed later in this section. The elements, which are covariant under left/right gauge transformations\footnote{The term $e^{\vartheta\phi}f$ is covariant under left gauge transformations. The terms $e^{\vartheta^\prime i}F_{u(1)}$ and $e^{\vartheta^{\prime\prime}i}F_{su(4)}$ are covariant under right gauge transformations.}, are formally assigned to {\it two sets of Clifford algebras}. The elements from different sets {\it formally commute with each other}. In the following, $\left\langle \cdots\right\rangle$ means the scalar part of both sets.
 
Before the final gauge action is given later in this section, a tentative gauge- and diffeomorphism-invariant gauge action is written as
\begin{equation}
S_G = S_{G1} + S_{G2} + S_{G3} + \cdots,
\end{equation}
where
\begin{align}
&S_{G1} \sim \int{\left\langle\mathcal{F}\wedge \mathcal{F}\right\rangle}, \\
&S_{G2} \sim \int{ \frac{\left\langle(\mathcal{F}\wedge \mathcal{F})^2\right\rangle}{\left\langle \eta\right\rangle}}, \\
&S_{G3} \sim \int{ \frac{\left\langle(\mathcal{F}\wedge \mathcal{F})^3\right\rangle}{\left\langle \eta^2\right\rangle}}, \\
&\cdots.\nonumber 
\end{align}
The $4$-form $\eta$ is defined as
\begin{equation}
\eta \equiv \phi^{\prime} D\phi^{\prime}\wedge D\phi^{\prime}\wedge D\phi^{\prime}\wedge D\phi^{\prime}, 
\end{equation}
where the Higgs field $\phi^{\prime}$ transforms in the same way as the Higgs field $\phi$, and it acquires a VEV $\bar{\phi}^{\prime}=v\bar{\phi}=vi$ with $v$ being a real constant. Here, we have assumed that $\eta$ is non-degenerate. It's understood that the parameter set $\{g, g^\prime, g^{\prime\prime}, \vartheta, \vartheta^\prime, \vartheta^{\prime\prime} \}$ may not  be the same for each $\mathcal{F}$. In the following, we differentiate the parameters by the subscript $k = 1, 2, \ldots $.

For further study of the action in the phase of broken symmetry, the Higgs fields are replaced by their VEVs, and connection fields $\omega,(1/l)e, W, A_{u(1)}, A_{su(4)}$ only are considered. The fields $W, A_{u(1)}, A_{su(4)}$ formally commute with the de Sitter connection $\omega + (1/l)e$. All the surviving symmetries after symmetry breaking are related to either part or combination of these gauge fields. The connection fields left out are suppressed by the Higgs mechanism. The curvature $2$-form $f$ can be written as
\begin{align}
f &= d(\omega + \frac{1}{l}e + W) + (\omega + \frac{1}{l}e + W) \wedge (\omega + \frac{1}{l}e + W)\nonumber \\
	&= R + \frac{1}{l^2}e\wedge e + \frac{1}{l}T + F_{W}
\end{align}
where the {\it spin connection curvature} $R$, the {\it torsion} $T$, and the {\it weak interaction curvature} $F_W$ are defined by $R \equiv d\omega + \omega \wedge \omega$, $T \equiv de + \omega \wedge e + e \wedge \omega$, and $F_W \equiv dW + W \wedge W$, respectively. There is no cross-term between the weak interaction connection $W$ and de Sitter connections $\omega + (1/l)e$, because they commute with each other.

The action $S_{G1}$ is calculated as
\begin{align}
S_{G1} \sim 
	&\int{\frac{\sin(\vartheta_1 + \vartheta_2)}{g_1g_2}
	\left\langle R \wedge R i \right\rangle} \\
+	&\int{\frac{\sin(\vartheta_1 + \vartheta_2)}{g_1g_2}
	\left\langle \frac{1}{l^2}(e \wedge e \wedge R + R \wedge e \wedge e) i \right\rangle} \label{EC}\\
+ &\int{\frac{\sin(\vartheta_1 + \vartheta_2)}{g_1g_2}
	\left\langle \frac{1}{l^4}e \wedge e \wedge e \wedge e	i \right\rangle} \label{CC}\\
+	&\int{\frac{\cos(\vartheta_1 + \vartheta_2)}{g_1g_2}
	\left\langle R \wedge R  \right\rangle} \label{PONT1}\\
+	&\int{\frac{\cos(\vartheta_1 - \vartheta_2)}{g_1g_2}
	\left\langle \frac{1}{l^2}(e \wedge e \wedge R + R \wedge e \wedge e + T \wedge T) \right\rangle} \label{NY}\\
+	&\int{\frac{-2\sin(\vartheta_1)\sin(\vartheta_2)}{g_1g_2}
	\left\langle \frac{1}{l^2} (e \wedge e \wedge R + R \wedge e \wedge e) \right\rangle} \label{IMMI}\\
+	&\int{\frac{\cos(\vartheta_1 + \vartheta_2)}{g_1g_2}
	\left\langle F_W \wedge F_W  \right\rangle} \label{PONT3}\\
+	&\int{\frac{\cos(\vartheta^{\prime}_1 + \vartheta^{\prime}_2)}{g^{\prime}_1g^{\prime}_2}
	\left\langle F_{u(1)} \wedge F_{u(1)}  \right\rangle} \label{PONT4}\\
+	&\int{\frac{\cos(\vartheta^{\prime\prime}_1 + 						\vartheta^{\prime\prime}_2)}{g^{\prime\prime}_1g^{\prime\prime}_2}
	\left\langle F_{su(4)} \wedge F_{su(4)}  \right\rangle}, \label{PONT5}
\end{align}
where the pseudoscalar $i$ is from the left gauge algebra. There are no cross-terms between different curvatures, because these cross-terms have no scalar part of the Clifford algebra. The first three terms correspond to the MacDowell-Mansouri formulation \cite{DESI} of gravity\footnote{There are variations of MacDowell-Mansouri gravity in the context of the Clifford algebra $C\!\ell_{3,1}$ \cite{ADS} or $SO(5) BF$ theory \cite{BF1, BF2}. The geometrical meaning of MacDowell-Mansouri gravity is discussed in the context of the Cartan geometry \cite{WI}.}. The first term is a topological invariant of the Euler class (also called the Gauss-Bonnet term). The second term is the {\it Einstein-Cartan action} of General Relativity. The third term is the {\it cosmological constant term}. The terms \eqref{PONT1}, \eqref{PONT3}, \eqref{PONT4}, \eqref{PONT5} are topological invariants of the Pontryagin class. The term \eqref{NY} is a topological invariant of the Nieh-Yan class \cite{NIEH}. The remaining term \eqref{IMMI} is the {\it Holst term} \cite{HOLS}. Even if this term is not topological, it does not affect the classical equation of motion when the torsion $T$ is zero. However, the Holst term plays a significant role in the context of loop quantum gravity \cite{LOOP}. 

For the action $S_{G2}$, there is a factor of $v^{-5}$ coming from the Higgs field VEV $\bar{\phi}^{\prime}=vi$. In the limit of $v^{-5} \ll 1$ and small curvatures, we would be only interested in the terms that do not appear in $S_{G1}$ and contain products of two or less curvatures $R, T, F_W, F_{u(1)}$, and $F_{su(4)}$ (the term $(1/l^2)e \wedge e$ is not counted as a small curvature due to non-degenerate VEV of the vierbein field). They are listed as
\begin{align}
	&\int{\frac{1}{g_1g_2g_3g_4\left\langle e \wedge e \wedge e \wedge ei\right\rangle}
	\left\langle (e \wedge e \wedge R + R \wedge e \wedge e)^2 i \right\rangle}, \\
 &\int{\frac{1}{g_1g_2g_3g_4\left\langle e \wedge e \wedge e \wedge ei\right\rangle}
	\left\langle (e \wedge e \wedge R + R \wedge e \wedge e)^2 \right\rangle}, \\
	&\int{\frac{1}{g_1g_2g_3g_4\left\langle e \wedge e \wedge e \wedge ei\right\rangle}
	\left\langle (e \wedge e \wedge T + T \wedge e \wedge e)^2 i \right\rangle}, \\
 &\int{\frac{1}{g_1g_2g_3g_4\left\langle e \wedge e \wedge e \wedge ei\right\rangle}
	\left\langle (e \wedge e \wedge T + T \wedge e \wedge e)^2 \right\rangle}, \\	
	&\int{\frac{1}{g_1g_2g_3g_4\left\langle e \wedge e \wedge e \wedge ei\right\rangle}
	\left\langle (e \wedge e \wedge F_W + F_W \wedge e \wedge e)^2 i \right\rangle}, \label{THET1}\\
 &\int{\frac{1}{g_1g_2g_3g_4\left\langle e \wedge e \wedge e \wedge ei\right\rangle}
	\left\langle (e \wedge e \wedge F_W + F_W \wedge e \wedge e)^2 \right\rangle}, \label{YM1}\\
	&\int{\frac{1}{g_1g_2^{\prime}g_3g_4^{\prime}\left\langle e \wedge e \wedge e \wedge ei\right\rangle}
	\left\langle (e \wedge e \wedge F_{u(1)} + F_{u(1)} \wedge e \wedge e)^2 i \right\rangle}, \label{THET2}\\
 &\int{\frac{1}{g_1g_2^{\prime}g_3g_4^{\prime}\left\langle e \wedge e \wedge e \wedge ei\right\rangle}
	\left\langle (e \wedge e \wedge F_{u(1)} + F_{u(1)} \wedge e \wedge e)^2 \right\rangle}, \label{YM2}\\
	&\int{\frac{1}{g_1g_2^{\prime\prime}g_3g_4^{\prime\prime}\left\langle e \wedge e \wedge e \wedge ei\right\rangle}
	\left\langle (e \wedge e \wedge F_{su(4)} + F_{su(4)} \wedge e \wedge e)^2 i \right\rangle}, \label{THET3}\\
 &\int{\frac{1}{g_1g_2^{\prime\prime}g_3g_4^{\prime\prime}\left\langle e \wedge e \wedge e \wedge ei\right\rangle}
	\left\langle (e \wedge e \wedge F_{su(4)} + F_{su(4)} \wedge e \wedge e)^2 \right\rangle}, \label{YM3}
\end{align}
where the pseudoscalar $i$ is from the left gauge algebra. For the sake of brevity, the $\vartheta, \vartheta^{\prime},$ and $\vartheta^{\prime\prime}$ dependence and the factor of $1/(16v^5)$ are not shown. The first four terms are {\it higher-derivative gravity terms} including torsion \cite{EFFE, HIGH}. 
For flat spacetime, with the substitution of the vierbein with its VEV $\bar{e} = \delta^{I}_{\mu}\gamma_I dx^{\mu} = \gamma_{\mu} dx^{\mu}$, the terms \eqref{THET1}, \eqref{THET2}, \eqref{THET3} (the {\it theta terms}) reduce to topological invariants of the Pontryagin class, and the terms \eqref{YM1}, \eqref{YM2}, \eqref{YM3} reduce to the {\it Yang-Mills actions}. 

In the limit of $v^{-5} \ll 1$ and small curvature, contributions from the actions $S_{G3}, S_{G4},$ and so on are negligible compared with those from $S_{G1}$ and $S_{G2}$.

The additional pseudoscalar $i$ makes all the difference between the Einstein-Cartan action \eqref{EC} and the Holst term \eqref{IMMI}. The same is true between the theta terms \eqref{THET1}, \eqref{THET2}, \eqref{THET3} and the Yang-Mills actions \eqref{YM1}, \eqref{YM2}, \eqref{YM3}. The similarities between these two cases are more salient here than the conventional formulation \cite{GOP}. 

The Yang-Mills interactions for the connection fields $W$ and $A_{su(4)}$ are stronger than gravity, due to the fact that the terms in $S_{G1}$ related to the Yang-Mills fields are all {\it topological}, thus do not contribute to the classical equation of motion. In addition to $1/g^{2}$ or $ 1/(g^{\prime\prime})^{2}$, there is a factor of $1/(v^{5}g^{2})$ for the Yang-Mills actions in $S_{G2}$, which makes the effective Yang-Mills coupling constants appear to be large. However, the feebleness of the $U(1)$ interaction as well as the smallness of the cosmological constant and theta terms\footnote{The theta term corresponding to the strong interaction is extremely small. Different explanations, such as the Peccei-Quinn mechanism \cite{PQ}, have been proposed.} can not be naturally explained in this scenario, even though the first one can technically be addressed by setting $g^{\prime} \ll g, g^{\prime\prime}$. In this regard, we propose the final formulation of the gauge actions
\begin{align}
&S_{G1} \sim \int{\left\langle \mathcal{F}^{\prime}\wedge D\phi^{\prime}\wedge D\phi^{\prime} 
+ D\phi^{\prime}\wedge D\phi^{\prime} \wedge \mathcal{F}^{\prime}\right\rangle}, \label{G1} \\
&S_{G2} \sim \int{ \frac{\left\langle(\mathcal{F}^{\prime}\wedge D\phi^{\prime}\wedge D\phi^{\prime} 
+ D\phi^{\prime}\wedge D\phi^{\prime} \wedge \mathcal{F}^{\prime})^2\right\rangle}{\left\langle \eta\right\rangle}}, \\
&S_{G3} \sim \int{ \frac{\left\langle(\mathcal{F}^{\prime}\wedge D\phi^{\prime}\wedge D\phi^{\prime}
+ D\phi^{\prime}\wedge D\phi^{\prime} \wedge \mathcal{F}^{\prime})^3\right\rangle}{\left\langle \eta^2\right\rangle}}, \\
&\cdots,\nonumber 
\end{align}
where $\mathcal{F}^{\prime} = (1 + c\phi^{\prime})\mathcal{F}$, and c is a real constant. In this formulation, there is no topological terms in \eqref{G1}. However, there are the same kinds of Einstein-Cartan action(from $S_{G1}$, along with the Holst term) \eqref{EC} \eqref{IMMI}, cosmological constant term (from $S_{G1}$) \eqref{CC}, Yang-Mills actions (from $S_{G2}$) \eqref{YM1} \eqref{YM2} \eqref{YM3}, and theta terms (from $S_{G2}$) \eqref{THET1} \eqref{THET2} \eqref{THET3}, only with different coefficients. These coefficients are of the order\footnote{The coefficients are set to be dimensionless with the help of the characteristic curvature scale $R_c$. It is assumed that the dimensionless constants except $v$ in $\mathcal{F}^{\prime}$ are of order 1.}
\begin{align}
&C_{EC} \sim \frac{v^3}{l^2 R_c}, \\
&C_{CC} \sim \frac{v^3}{l^4 R_c^2}, \\
&C_{YM} \sim v, \\
&C_{\theta} \sim 1,
\end{align}
respectively. The feebleness of gravity as well as the smallness of the cosmological constant term (compared with the Einstein-Cartan action) and the theta terms (compared with the Yang-Mills actions) are the consequences if the conditions
\begin{equation}
v^2 \gg l^2 R_c \gg 1
\end{equation}
are satisfied. The parameters $v$ and $l$ are determined by the VEVs of the Higgs field $\phi^{\prime}$ and the gauge field $a_-$, respectively. It is understood that the above discussion takes place in a purely classical context. The quantum corrections and renormalizability (or rather non-renormalizability) are not considered here. 

The gauge fields we have left out (gauge fields in $a$ other than $\omega, (1/l)e, W$) are related to the broken gauge symmetries. They are not related to the residual symmetries \eqref{LO}, \eqref{SU3}, \eqref{EM}, or \eqref{U1} after the Higgs fields acquire their VEVs. These gauge fields are assumed to be suppressed via the unknown Higgs sector by acquiring large masses. However, at extremely high energy levels above the gauge field mass scales, these gauge fields could be detectable. And they would contribute to terms different from Einstein-Cartan and Yang-Mills actions, because these fields do not commute with the de Sitter algebra. The mixing of these gauge fields with the de Sitter connection may have interesting consequences under extreme conditions in the early universe or at the Large Hadron Collider. We leave the study of these effects to future research.


\section{Conclusion}

We present a model of Yang-Mills interactions and gravity in terms of the Clifford algebra $C\!\ell_{0,6}$, which has the same degree of freedom as that of real fermion components in the first generation. The electrons, neutrinos, and quarks of three colors are identified with even and odd parts of spinor idempotent projections. An intriguing feature is that if the leptons and quarks are labeled by ``+'' and ``-'' signs of the idempotents $p_{\pm{1}}, p_{\pm{2}}, p_{\pm{3}}, P_{\pm}$, and the chirality of the spinor, the assignments bear some resemblance to the 5-bit spinor representation of $SO(10)$ grand unified theory.

The invariance groups, including the de Sitter and the Pati-Salam $SU(4)$ subgroups, consist of gauge transformations from either side of an algebraic spinor. Upon symmetry breaking via the Higgs fields, the remaining symmetries are the Lorentz $SO(1,3)$, color $SU(3)$, electromagnetic $U(1)_{EM}$, and an additional $U(1)^\prime$. 

The gravity and Yang-Mills actions are formulated as different order terms in a generalized action. The first order terms include the gravity action. The effective Yang-Mills coupling constants seem to be large, because the Yang-Mills actions are part of the second order terms. The feebleness of gravity as well as the smallness of the cosmological constant and theta terms are discussed at the classical level, with the VEV magnetitudes of the Higgs field $\phi^{\prime}$ and the gauge field $a_-$ playing an essential role.

There are still several shortcomings with the current model. The major issue is that the Higgs sector is not provided. Without the Higgs mechanism of spontaneous symmetry breaking\footnote{See ref. \cite{STE, LEC} for attempts to construct a Higgs sector for gravity.}, neither do we know how the Higgs fields acquire the rather {\it ad hoc} VEVs, nor do we know how the gauge bosons with broken symmetries are suppressed by acquiring large masses. 

The second issue is the difficulty of explaining the chirality of weak interactions, since there are only three degrees of freedom for the weak transformations applying to both the left- and right-handed components of the spinor. Another issue is to find a mechanism to suppress the fifth force related to the $U(1)^\prime$ symmetry. Further research is needed to fully recover the standard model results.

\section*{Acknowledgments}

I am grateful to Chris Doran, Muyu Guo, David Hestenes, and Lee Smolin for helpful comments on the drafts of this paper.

\end{document}